# Using Four Different Online Media Sources to Forecast the Crude Oil Price

Elshendy, M., Fronzetti Colladon, A., Battistoni, E. & Gloor, P. A.

# Using Four Different Online Media Sources to Forecast the Crude Oil Price

**Mohammed Elshendy**
Department of Enterprise Engineering, University of Rome Tor Vergata, Italy.

**Andrea Fronzetti Colladon**
Department of Enterprise Engineering, University of Rome Tor Vergata, Italy.

**Elisa Battistoni**
Department of Enterprise Engineering, University of Rome Tor Vergata, Italy.

**Peter A. Gloor**
MIT Center for Collective Intelligence, Massachusetts Institute of Technology, US.

**Abstract**
This study looks for signals of economic awareness on online social media and tests their significance in economic predictions. The study analyses, over a period of two years, the relationship between the West Texas Intermediate daily crude oil price and multiple predictors extracted from Twitter, Google Trends, Wikipedia, and the Global Data on Events, Language, and Tone database (GDELT). Semantic Analysis is applied to study the sentiment, emotionality and complexity of the language used. ARIMAX models are used to make predictions and to confirm the value of the study variables. Results show that the combined analysis of the four media platforms carries valuable information in making financial forecasting. Twitter language complexity, GDELT number of articles and Wikipedia page reads have the highest predictive power. The study also allows a comparison of the different fore-sighting abilities of each platform, in terms of how many days ahead a platform can predict a price movement before it happens. In comparison to previous work, more media sources, and more dimensions of the interaction and of the language used, are combined in a joint analysis.

## 1. Introduction

The art of prediction has a long history. People have always tried to design schemes, implement hypotheses and sometimes create legends to forecast certain outcomes of physical and natural events. In financial market studies, the main aspect of making correct predictions depends on identifying the most relevant and critical predictors. Some of these are sometimes difficult to identify or measure, are constantly changing, or may not have been completely explored.

Surowiecki's [1] seminal book "The Wisdom of Crowds" claims that large groups of individuals are better at making decisions about uncertain events than experts. The wisdom of crowds is meant to be the public opinion or mood. In terms of gathering, quantifying and qualifying wisdom of crowds, business executives and Academia have found an instrument in the social media.

Apparently, social media has exploded as a category of online discourse where people create, share, bookmark and network contents at a prodigious rate [2]. According to "We Are Social" – a comprehensive study of digital, social and mobile usage around the world (http://wearesocial.com/uk/special-reports/digital-in-2016) – there were 3.419 billion internet users at the start of 2016, representing almost 46% of the Earth's population. Out of these, there were 2.307 billion people – 31% of the global population – who were active users of social media platforms. It is therefore reasonable to say that social media represents a revolutionary trend, being always more of interest to companies operating in online space or in any space [3].

This paper tests the relevance of selected online open data sources to forecast financial events. It analyses in depth the daily traffic activity of four different social media platforms and tools, in order to predict the West Texas

Intermediate (WTI) Crude Oil Price. In this context, information from the following online platforms were integrated: (a) the microblogging and media-sharing platform – Twitter; (b) the user generated web encyclopaedia – Wikipedia; (c) the online traffic analysis platform – Google Trends; and (d) the world news database of the GDELT Project. The choice of predicting the WTI Crude Oil Price is due to its chaotic behaviour, which follows a nonlinear dynamic deterministic process – as proved by Moshiri and Foroutan [4]. Thus, understanding the dynamics of Crude Oil Prices, either spot or futures, represents one of the most striking challenges to the forecasting abilities of private and public institutions worldwide [5].

While making the predictions that have just been mentioned, one should consider that many forces interact in the oil industry and can ultimately affect the crude oil price. Past research about the determinants of the oil price can be divided into two main streams: the former considering oil supply, demand and the marketing chain, focused on the analysis of the desires of buyers and on the production capacity of suppliers [6,7]; such elements are in many cases affected by geo-political agreements and commercial choices of the OPEC members [8]. For example, Chevillon and Rifllart [9] stated that crude oil prices are mainly affected by agents involved in the production, processing, and merchandising of the oil commodity in private or governmental sectors. A second stream of research is instead more focused on the market sentiment and on the general attitude and psychology of investors [10,11]; these elements can be influenced by the behaviour of financial experts and speculators, by the online dialogue on microblogging platforms such as Twitter, and by the news stream. Yet this study was not conceived with the idea to prove this influence, or to carry out an in-depth analysis about the oil industry and the determinants of the oil price, but to discuss the predictive power of four online media platforms where the discourse about the oil price is reflected, as well as the discourse about potentially connected events – such as wars or oil shortages. The authors aim to help the analysts who are used to consider all the above mentioned forces, with the introduction of new indicators as potentially useful addition to existing forecasting models.

## 2. Literature Review: Predicting the Oil Price

Oil price forecasting techniques can be classified into two main categories: econometric/financial methods and computational models. Econometric models are usually based on a quantitative approach, combining oil price historical data with analytical models. Scholars make frequent use of ARIMA-, ARCH/GARCH-, and Markov Switching-techniques e.g. [12]; in several cases these models do not include external predictors and just rely on the historical trends of the input variable, or on few selected control variables, such as the Nasdaq 100 and the SP500. On the other hand, computational models are mostly based on the inclusion of explanatory variables, either oil price or economic related, in order to explain the relationship between spot and future prices. They make a frequent use of Artificial Neural Networks (ANN), Support Vector Machine (SVM) and text mining techniques. Recently, the use of computational models is more frequent due to the major advances of computer technologies and the large pools of data available online [13,14]. Replicating these models can however be more complicated, since their setup is mostly data dependent and the contribution of single features is sometimes more difficult to isolate. Since one of the scopes of this research is to show the predictive potential of four different media platforms, the choice has been to use ARIMAX models which can include external independent variables and, at the same time, give evidence of the contribution of each of them. In addition, the findings described in the paper can serve as the base for future predictive models which use a machine learning approach.

Cheong [15] proposed an ARCH model to forecast the time varying volatility of crude oil prices on the short, medium and long period for the WTI and Europe Brent benchmarks. Similarly, Wei et al. [16] proposed nine linear and nonlinear generalized GARCH models to capture the volatility features for the same benchmarks. Results showed that the linear models were useful for volatility predictions but only on the short run (i.e. one day). By contrast, the nonlinear models exhibited higher forecasting accuracy, especially on longer time intervals (i.e. five to twenty days).

Fernandez [17] proposed a non-seasonal ARIMA model for the time series analysis of daily prices of the Dubai crude oil and natural gas. Their models were based on performing sample autocorrelation and partial autocorrelation tests. Results showed the superiority of the model in making short-term predictions even when compared to ANN and SVM models; by contrast, SVM models proved to be the best when making prediction in the long term. Bosler [18] compared the prediction performance of the same three models, applying the analysis to quarterly data of free on board (FOB) crude oil prices. Partially contrasting [17], results showed a worse performance of the SVM approach.

Shambora and Rossiter [19] forecasted crude oil price by building an ANN model and analysing the traded futures contracts on the period between April 1991 and December 1997. In addition, they applied a random walk model and a simple moving average model. Empirical results showed that the ANN model is more accurate than the others. However, the overall performance of the three models suggested that traded futures contracts are not an efficient predictor.

Yu et al. [20] proposed a Refined Text Mining approach for crude oil price forecasting. The model was based on the refining of a dataset obtained extracting unstructured text documents, related to the crude oil price query, from Google results. They compared the forecasting ability of their model with regression, random walk, and ARIMA models. They concluded that random walk and regression models performed worst when compared to ARIMA and refined text mining. Wang et al. [21] proposed a novel nonlinear integrated model called TEI@I to predict WTI Crude Oil Prices: they integrated ARIMA and a back propagation neural networks. In this way, they were able to capture both linearity and non-linearity characteristics of the time series. In addition, they investigated the existence of irregular activities during the studied period by using web-based text mining techniques, reaching good overall performance results.

Similarly, this study combines ARIMAX time series models with text mining. Therefore, an integrated model combining both econometric and computational techniques is proposed. The historical data of crude oil price are used to explain spot prices, in addition to the inclusion of 10 external independent variables. These variables are non-economic; however, they track the online discourse about crude oil price and include world broadcasts, print, and web news. To the authors' knowledge, there are no previous studies trying to forecast the oil price from the same sources, taken all together. However, some of the variables used here were also considered in other financial predictions: they are discussed in the coming section.

## 2.1. Hypotheses formulation

In this study, the authors try to forecast the crude oil price, exploring the data available on four different online media platforms. The research questions they try to address are the following: with which accuracy is it possible to predict the oil price using the data available on Twitter, Google Trends, Wikipedia and GELT? Does the integration of more platforms provide better outcomes? Do these platforms offer valuable predictors (that could eventually represent a significant addition to other existing forecasting models)? To answer these questions, authors examined the research related to each media platform and developed new predictive models. Since, to the best of the authors' knowledge, some data platforms, like GDELT, have not been used yet to predict the crude oil price, this study includes some novel elements.

In an early effort in using Twitter for financial predictions, Zhang et al. [22] collected six months of Twitter feeds, analysing the overall emotional state of Twitter users who expressed their optimistic and pessimistic thoughts about gold price, crude oil price, currency exchange rates, and stock market prices, on a daily basis. The study showed a significant correlation between the crowd's emotions on Twitter and all the dependent variables, except for the US dollar exchange rate. Similarly, Bollen et al. [23], Mittal and Goel [24], Chen and Lazer [25], Sprenger et al. [26] applied further studies on stock market indicators using volume and sentiment of Twitter feeds, showing that Twitter feeds are effective indicators of real world performance. In addition to Twitter feeds and sentiment, Rao and Srivastava [27] included the search volume index from Google Insights of Search (currently known as Google Trends), in an effort to model and predict oil, gold, and forex market indices. Results showed that their proposed model outperformed earlier works in terms of accuracy and forecasting errors, even if taking into consideration only Twitter sentiment and feeds. Additionally, Twitter prediction ability has been studied in several fields and has shown a good degree of success, e.g. political elections [28, 29], box office [30,31], or sales [32,33]. These studies showed significant correlations of Twitter feeds (i.e. number of tweets) and Twitter sentiment with their respective dependent variables. In the financial field, a higher volume of tweets has been usually related with a state of concern, which can create a large discussion, usually immediately after a financial event. Thus, it more frequently leads to price drops. Oh et al. [34] showed how the Twitter traffic dramatically increased after 2008 Mumbai terrorist attacks, which produced major loss in stock returns [35]. On the other hand, a positive sentiment of the language used represents a state of optimism, which can positively reflect on prices [36]. Accordingly, the following hypotheses are formulated:

H1a: The daily count of tweets is negatively related with oil price.
H1b: The more positive the sentiment of the language of the tweets, the higher the oil price.

Twitter analysis was extended to consider other two dimensions of the language used, which are usually less explored. Firstly, language emotionality, which is calculated as the deviation from neutral sentiment; secondly, the language complexity, which assigns a complexity level to each tweet by measuring how much each term is commonly used in the overall discourse [37]. Forbergskog [38] investigated the leading and lagged relationships between positive and negative emotions on the Twitter feed regarding the S&P 500 index, over 84 trading days. Porshnev et al. [39] investigated the possibility of analysing the emoticons in the tweets with the aim of forecasting DJIA and S&P500 stock market indices. Both studies concluded that an increase of emotionality in tweets – conceived in the present study as a larger variation between positive and negative sentiment – leads to a reduction of stock market performance and price drops. On the other hand, there seem to be no available studies linking language complexity to financial prediction. Authors hypothesise that a higher complexity is related with price drops, as the language becomes more heterogeneous and less shared, which can be connected to new radical events or to the intervention of specialized Twitter actors – experts, professional traders or journalists – commenting on such, mostly negative, events [40].

> H1c: High emotionality values of twitter feeds are negatively related with oil price.
> H1d: High complexity values of twitter feeds are negatively related with oil price.

Used broadly in the economic and financial fields, Google Trends showed wide tracking and real time surveillance abilities. For instance, it was used for now-casting of macroeconomic content and probabilities [41,42], or of unemployment rates [43-45]. Wu and Brynjolfsson [46] conducted an explanatory study on how Google search engine data provides an accurate and simple way to predict housing prices and sales activities. Results showed that the housing search index is strongly predictive of future housing markets and sales. Fantazzini and Toktamysova [47] studied the effect of adding Google Trends to other economic variables in forecasting German car sales. Results showed that new models statistically outperformed old models (not considering Google Trends) for most of the automobile manufacturers and forecast horizons. Choi and Varian [48-50] described how to use Google Trends data to predict several economic metrics, including unemployment rates, automobile demand, and vacation destinations. In their 2012 report [50], they presented short-term forecasts of multiple economic indicators, showing that the inclusion of Google Trends in the analysis could improve model outcomes by 5% to 20%. Their examples showed a positive association of the volume of search queries with the financial and economic indicators. Moreover, Scott and Varian [51] predicted gun sales at the national level by analysing 100 different queries related to the subject. Consistently, this study includes Google Trends search activity in the forecasting of oil price, considering two search queries ("Price of Oil" and "OPEC"). Other queries were also tested with no better associations. The expectation is to find a positive link between the search activity and the oil price.

> H2: The query counts on Google Trends for terms "Price of Oil" and "OPEC" are positively related with oil price. More counts means higher prices.

Recently, Wikipedia has attracted a big attention from scholars working in several fields. For instance, Gloor et al. [52] analysed Wikipedia networks of world leaders and introduced a dynamic temporal map of the most influential people of all time. Yasseri and Bright [53] explored the use of Wikipedia page view data in electoral predictions, for the elections of the members of the European Parliament in 2009 and 2014.

Mestyán et al. [54] proposed a predictive model for the financial success of movies (before their release) based on the activity level related to the movie in Wikipedia. Predictive power was strong a few days before the release. Wei and Wang [55] proved the existence of a link between the number of page views of a company in Wikipedia and its subsequent performance in the stock market. Moat et al. [56] investigated whether the historical data of Wikipedia page views may contain signs of stock market movements. They considered the page views and edits for all the companies listed in the Dow Jones index, stressing the importance of investigating the activity on Wikipedia while making financial predictions, and finding a negative association of page views with stock prices. With regard to the oil price, similarly to Twitter, a large number of page views could reflect a larger state of concern coming after significant events affecting the financial world. Accordingly, the following hypothesis is formulated:

> H3: The page views count of the specialized Wikipedia articles "Price of Oil" and "OPEC" are negatively related with oil price.

As regards the more recent GDELT project, scholars have not yet fully explored the link between its data and the trends on financial markets. Up to date, research efforts are mostly focused on political conflicts and demonstrations. Yonamine [57] used the dataset to predict future levels of violence in Afghanistan districts. Kwak and An [58] studied GDELT for understanding news geography and major determinants of global news coverage of disasters. Phua et al. [59] conducted a visual and predictive analysis of Singaporean news coverage. They compared Singaporean news articles from 1979 to 2013 with the Wikipedia timeline of the Singaporean history; they used a machine learning algorithm to extract and match the main topics emerging from the two platforms. Subsequently, they used the GDELT data to predict the Singapore stock market's Straits Times Index. Bodas-Sagi and Labeaga [60] used GDELT to analyse the public opinion about the energy policy of the Spanish Government. They carried out the analysis twice, the first time collecting all the data relating to the query "Fuel Price", and then adding only governmental data relating to the same query. Both cases showed a negative association between GDELT activity and Spanish Government energy policy. In this study, a similar trend is expected when querying the database for "WTI Crude Oil Price" related news. Indeed, there is a possible connection between the news activity related to the energy policies and the crude oil price. Moreover, the expectation of a negative association between the oil price and the GDELT activity is consistent with the hypotheses formulated for Twitter and Wikipedia. By contrast, the number of organizations mentioned with regard to the WTI Oil Price (in the news on GDELT) is expected to have a positive association with the oil price: enterprises are often linked to new projects, positive events or business agreements.

>    H4a: A higher number of articles on GDELT, for the search query "WTI Crude Oil Price", is negatively related with oil price.
>    H4b: The number of organizations names mentioned on GDELT, with regard to the search query "WTI Crude Oil Price", is positively related with the oil price.

## 3. Methodology

The daily spot price of the WTI Crude Oil has been analysed from April 1, 2013 to April 1, 2015. The choice of the WTI spot price has been made because of its availability and its common usage in the literature of oil price modelling and prediction. Price data have been obtained from the U.S. Energy Information Administration. In addition, ten independent variables representing the online traffic of the four social media platforms – Twitter, Wikipedia, Google Trends and the GDELT Project – have been collected. The aim has been to study the "oil price" search term from different resources to target different publics.

### 3.1. Data Sources and Collection

The choice of the four media platforms included in the analysis was founded on their size – in terms of number of users/media sources and produced content (the bigger the data source, the more the information that can be analysed) –, by their reputation, and by the possibility to get free access to the data. Each specific choice is better supported in the following.

Twitter is a worldwide popular platform, which offers a social networking and microblogging service. It enables its users to update their status in tweets, to follow people they are interested in, and to communicate with them directly. Twitter users include, but are not limited to, commodity traders, politicians, companies, activists, and major news outlets, as well as casual users. Twitter describes itself as "a real-time information network that connects you to the latest information about what you find interesting". It is a social awareness system, which delivers a fragmented mix of information, enlightenment, entertainment, and engagement from a range of sources [61]. In 2012, Twitter had more than 100 million registered users posting more than 340 million tweets per day[1], updated to be 310 million monthly active users on March 2016[2]. Thus, Twitter popularity has drawn more and more researchers' attention from different disciplines, to understand its usage and community structure, influence of users and information propagation, and its prediction power and potential application to other areas [22]. Twitter data have been collected using the social network and semantic analysis software Condor, developed by Galaxy Advisors[3]. It enables the visualisation, measurement, and analysis of the communication structure in social networks, as well as the analysis of the use of language over time.

We collected data from Twitter by fetching all the tweets that contained the search term "crude oil price". Collected tweets were then filtered, excluding all non-relating posts – for instance, those referring to cooking or lubricating oils. Analysing the tweets, four Twitter variables have been determined, i.e. Number of tweets per day, Sentiment, Complexity and Emotionality.

In addition to Twitter, Wikipedia, the well-known online encyclopaedia, has been included in this research. Currently, Wikipedia is the largest knowledge repository on the web. Wikipedia is available in dozens of languages, and its English version is the largest of all with more than 400 million words in over one million articles [62]. It is also densely structured: its articles have in total hundreds of millions of links. These connections link the topics being discussed, and provide an environment which fosters serendipitous gathering of information [63]. This huge amount of information and links provides a real opportunity to help unfold the world history and explore upcoming events. Different from Twitter, the quality of information in Wikipedia is controlled and consequently higher. Within the Wikipedia research community, findings are constantly published and verified, and the reputation of an author grows with the reputation of his contributions [64]. The distribution of users' reputation in Wikipedia shows that saboteurs and inexpert users are quite a minority compared to high reputation users [65].

For Wikipedia, the daily views count of the pages relating to Crude Oil Price, such as "Benchmark (crude oil)", "World oil market chronology", "2000s energy crisis", and others, were collected. For the final study, two pages have been considered – "Price of oil" and "OPEC" – as their traffic statistics were the most significant in terms of correlation with the dependent variable.

Google is the first search engine in the world, which makes it one of the most reliable resources for investigating web search queries. Since 2004, Google has been providing three data sources that can be useful for social science: Google Trends, Google Correlate, and Google Consumer Surveys [66]. Google Trends is commonly used in "now-casting" or in the prediction of the present, the very near future, and the very near past [67].

In this study, the search queries were set to be the same as the page titles of Wikipedia, mainly for two reasons: (1) to show how different public audiences can produce different predictive activity using the same keywords on two different platforms; (2) to show the response time difference between the two platforms. An analogous procedure to that of Wikipedia has been applied to gather Google Trends data. This data is available directly on the Google website, which allows filtering for location, time range, and keywords. The location has been set to "worldwide", whilst the queried keywords have been "Price of oil" and "OPEC". Other possible keyword combinations have been tested, obtaining less significant results.

The Global Data on Events, Location and Tone (GDELT) is the final platform considered. The GDELT Project is an open source repository of news articles, which is continuously updated and made available to researchers through an application program interface. Initially, the GDELT Project was a coded dataset of 200 million geo-located events; now, it has been updated to be 400 million events spanning over more than 12,900 days[4]. The dataset includes more than 300 different types of events: therefore it reveals all that has happened in place and time, since 1979. Kwak and An [68] referred to it as a tale of the world. The GDELT Project is described as an initiative to construct a catalogue of human societal-scale behaviour and beliefs across all countries of the world. It connects every person, organization, location, count, theme, news source, and event across the planet into a single massive network. The database relies on tens of thousands of broadcasts, print and online news sources from every corner of the globe [69]. Including GDELT in the analysis is important to control for world events related to the WTI Crude Oil Price and to have a proxy for media activities on newspapers.

GDELT data was obtained by crawling the Global Knowledge Graph (GKG) dataset available on the project website. The daily number of newspaper articles covering "WTI Crude Oil Price" search criterion and the count of all organizations names or advisory councils mentioned all over the world during the period of the study have been extracted.

### 3.2. Description of Variables

A detailed description of all the independent variables is presented in Table 1.

**Table 1.** Description of independent variables.

| Platform | Variable | Description |
|---|---|---|
| Twitter | Messages per Day | It is the count of the number of tweets written in the English language, which contains the search query on a given day. |
| | Sentiment | Sentiment varies in the range [0,1], where a higher score represents a more positive language with respect to the query term. The calculus was made by means of machine learning algorithms included in the software Condor [70]. |
| | Emotionality | Emotionality is calculated as the deviation from neutral sentiment [70]. |
| | Complexity | Text complexity is measured considering the occurrence of single words in the whole text: less used words have higher complexity than the other commonly used words. We then calculate text complexity as suggested by [65], considering the Inverse Document Frequency (IDF), which reflects how important a word is to a document or a corpus. |
| Wikipedia | Page Visits | A count, which refers to how many times a certain article has been visited during a specific timeframe. |
| Google Trends | Query Count | A count which shows how often a particular keyword or query is searched in relation to the total search volume in a certain geographical area at a certain time period. |
| GDELT | Number of Articles | A numerical quantification, which measures the volume of news coverage nominating the "WTI Crude Oil Price" as the main event in a particular geographical location. |
| | Number of Organizations | Organizations mentioned in articles matching the searching criteria. It includes multiple organizations names and councils, expressing both non-profit and commercial enterprises. |

### 3.3. Data Analysis

As discussed in Section 2, a wide body of literature has studied the oil price prediction using various methods, including, but not limited to, time series analysis, artificial neural networks, support vector machines, empirical mode decomposition, and wavelet transforms, in addition to the traditional linear and non-linear regressions, error corrections and econometric methods. Many of these studies present some limitations. For instance, they often consider weekly or monthly data in time series forecasts e.g. [71,72], thus shrinking the dataset size, and not allowing a day by day prediction. Moreover, taking 300 observations on a monthly basis means going back 25 years, which means including in the study a different or an irrelevant economic situation compared to the present one. Furthermore, it limits the ability of dealing with daily lead/lag. In addition, several studies base their forecasts on a set of observations on a single variable e.g. [73]. Therefore, the output series is described in terms of past values of the input series. Forecasts from such models are therefore only extrapolations of the observed series [74].

In this context, the choice has been to implement both ARIMA and ARIMAX time series forecasting models, for two main reasons. First, when compared to other techniques – such as multiple linear regression models – time series models are usually superior in making daily forecasts, as they are built considering the autocorrelation function (ACF) and the partial autocorrelation function (PACF). Accordingly, including a differencing order, these models can achieve data stationary and uncover the presence of unit roots and trends [75]. Second, the ARIMA model produces forecasts based on past values in the time series (AR terms) and on the errors made by previous predictions (MA terms). Those parameters allow the model to auto-adjust efficiently upon sudden changes, resulting in higher accuracy. The ARIMAX procedure takes the ARIMA model one-step ahead, by including external factors or independent variables into the model [76].

One concern has been dealing with the time series gaps. Missing values during weekends and public holidays have been excluded, while missing data during working days due to shutdowns or unknown reasons have been interpolated, in order to create a five days per week observation time. The lags were limited to a maximum of three days: indeed, the missing data – either in the Crude Oil Price during weekends and holidays or in the platform variables due to failures – returns inaccurate prediction results on higher lags and causes computational failures in financial econometric models. The data set has been trained on the period from April 1, 2013 to March 14, 2015 as sample data. The period from March 15, 2015 to April 1, 2015 has been used to test the models on real predictions, i.e. has been considered as the out-of-sample. To evaluate and compare the accuracy of forecasts, the Root Mean Squared Error (RMSE) and the Mean Absolute Percentage Error (MAPE) of the predictive models have been calculated.

## 4. Results

This section discusses the prediction ability of each platform in terms of how many days the WTI Crude Oil Price takes to react to the platform activity. In statistical words, the lagging of independent variables is held to predict what will happen in time $t$, based on the knowledge available at the time $(t–n)$, where $n$ is the number of lags. For instance, to predict the price value on Friday, first, second, and third order lags have been created. The first lag dataset uses the data available on Thursday, the second lag dataset refers to data available on Wednesday, and the third lag dataset refers to data available on Tuesday. To further test the robustness of predictors, the study includes two control variables: the WTI Crude Oil Price of the previous days, and the NASDAQ 100 index. The choice of the NASDAQ 100 index over other well-known indexes – as the S&P 500 – was driven by the fact that this index has proved to be good in reflecting general changes in the stock market, without being too biased by the financial activities of a specific set of organizations [77].

## 4.1. Correlation analysis

Pearson correlation analysis was used to test the association between the social awareness variables and the WTI Crude Oil Price on the first, second and third lags. Results are shown in table 2.

**Table 2.** Correlations with Oil Price

| Sources | | One-day lag | Two-day lag | Three-day lag |
|---|---|---:|---:|---:|
| Control Variables | WTI Crude Oil Price | 0.967* | 0.938* | 0.910* |
| | Nasdaq 100 | -0.682* | -0.657* | -0.630* |
| Twitter | Twitter Messages | -0.508* | -0.515* | -0.521* |
| | Twitter Sentiment | 0.503* | 0.479* | 0.458* |
| | Twitter Emotionality | -0.097 | -0.090 | -0.085 |
| | Twitter Complexity | -0.280* | -0.245* | -0.222* |
| Google Trends | Google Trends "OPEC" Query Count | 0.617* | 0.618* | 0.617* |
| | Google Trends "Price of Oil" Query Count | 0.359* | 0.357* | 0.354* |
| Wikipedia | Wikipedia "OPEC" Page Views | -0.562* | -0.567* | -0.576* |
| | Wikipedia "Price of Oil" Page Views | -0.785* | -0.785* | -0.785* |
| GDELT | Number of Organizations | 0.267* | 0.274* | 0.279* |
| | Number of Articles | -0.717* | -0.720* | -0.716* |

N = 372, *p<0.05

From the results in Table 2 we can draw the following considerations:

(1) Twitter variables, except for Emotionality, are significantly related with the price. Sentiment and Complexity show a better correlation on the first lag, whilst the Number of tweets correlates the best at a three-day lag. It seems that a lower number of tweets, together with a higher Sentiment and a lover Complexity, are signals of higher oil prices.
(2) Google Trends and Wikipedia variables are also significantly correlated with the WTI Crude Oil Price, at all lags. The search query "Price of Oil" shows a higher correlation at the first lag for Google Trends, whilst the search query "OPEC" is higher on the second lag. These differences are relatively small, also for the Wikipedia variables. While the relation of oil price with the Wikipedia predictors is negative, the one with Google Trends is positive.
(3) The GDELT count of number of articles shows a much higher correlation to oil price than the count of the number of organizations. The number of articles shows a higher positive correlation on a two-day lag, whilst the number of organizations shows a higher negative correlation on a three-day lag.
(4) Finally the NASDAQ index shows negative correlation with the oil price. As one could expect, the oil price results are highly correlated with itself, at all three lags.

## 4.2. Granger causality tests and ARIMA/ARIMAX models

To check the predictive power of each of the independent variables, Granger causality analysis [78] was implemented. Granger causality is a statistical test, which reveals if one time series provides useful information which help in forecasting another time series (Table 3). The number of observations is 372 and it refers to the number of days included in the observation period. Per each day, thousands of entries coming from the different platforms have been analysed.

**Table 3.** Granger causality test results.

| Variable | First Lag ($\chi^2$) | Second Lag ($\chi^2$) | Third Lag ($\chi^2$) |
|---|---:|---:|---:|
| Nasdaq 100 | 12.69* | 9.605* | 8.702* |
| Twitter Messages | 1.016 | 1.306 | 1.959 |
| Twitter Sentiment | 0.339 | 0.296 | 2.380 |
| Twitter Emotionality | 9.920* | 8.452* | 7.061* |
| Twitter Complexity | 12.32* | 1.976 | 4.606 |
| Google Trends "OPEC" Query Count | 0.275 | 1.356 | 0.0315 |
| Google Trends "Price of Oil" Query Count | 6.851* | 8.824* | 12.522* |
| Wikipedia "OPEC" Page Views | 7.010* | 14.047* | 0.5586 |
| Wikipedia "Price of Oil" Page Views | 1.204 | 8.277* | 11.153* |
| GDELT Number of Organizations | 0.723 | 0.194 | 0.915 |
| GDELT Number of Articles | 3.076 | 4.355* | 5.348 |

N = 372; *p<0.05.

Similarly to ARIMA/ARIMAX, determining Granger causality requires the time series to be stationary. Since most of financial time series are subject to trends, seasonality and periodic variations, the stationarity hypothesis is probably rejected. In order to check the series stationarity, the augmented Dicky Fuller unit root test has been applied to all of the predictors, which have shown a non-stationary behaviour and have required a first order differencing to achieve stationarity.

Table 3 shows a significant Granger correlation of Crude Oil Price with Twitter Emotionality and Complexity, Google Trends "price of oil" queries count, Wikipedia "price of oil" and "OPEC" page views count, and GDELT number of articles. Moreover, the lag orders reveal a lot about each platform and its users. Twitter Complexity is the most predictive on a one-day lag and it correlates negatively to the WTI oil price. A less shared language – probably more technical and mostly used by specialized traders – can predict a price drop one day before it happens. Similarly, Twitter Emotionality – originally not correlated with the WTI Crude Oil Price – shows the largest predictive power at a one-day lag, after the differencing.

On the other hand, Google Trends and Wikipedia "Price of Oil" seem to require more time to be updated and are mostly effective at a three-day lag. The article title "OPEC" – which is a more specialized term in respect to "price of oil" – is only predictive for Wikipedia, at a one-day lag. It could happen that some traders consult the OPEC page on Wikipedia before choosing whether to buy or not. Finally, the GDELT count of number of articles shows prediction ability only on a two-day lag.

Starting from Table 3 outcomes, multivariate ARIMAX models have been trained. Having considered the autocorrelation function (ACF), partial auto correlation function (PACF), Akaike information criterion (AIC) and Bayesian information criterion (BIC) of the WTI Crude Oil Price time series, the ARIMA (2,1,4) model has revealed to offer the best parameters. Model outcomes have been evaluated twice. Once, by comparing the AIC and BIC measures for in-sample fitted values. The second time, by means of RMSE and of MAPE, based on the out-of-sample prediction. Table 4 shows the final results.

Models are presented in consistent variables blocks. The split of Twitter variables (Models 2 and 3) has been necessary due to collinearity problems.

The ARIMAX models outperform the ARIMA models regardless of the platform used in prediction. In terms of RMSE and MAPE, the best ARIMA model shows forecast errors of 13.878 and 22.579, respectively; in comparison, the best ARIMAX model shows errors of forecast of 1.683 and 2.498, respectively.

**Table 4.** Evaluation of ARIMA/ARIMAX (2,1,4) models.

| Variable | ARIMA | | | | | | | ARIMAX |
|---|---|---|---|---|---|---|---|---|
| | Model 1 | Model 2 | Model 3 | Model 4 | Model 5 | Model 6 | Model 7 | Model 8 |
| Nasdaq 100 (t–1) | | -0.001 | | | | | | -0.002 |
| Twitter Messages (t–3) | | | -0.001 | | | | | |
| Twitter Sentiment (t–3) | | | -1.481 | | | | | |
| Twitter Emotionality (t–1) | | | -7.524 | | | | | |
| Twitter Complexity (t–1) | | | | -0.689* | | | | -0.810* |
| Google Trends "OPEC" Query Count (t–2) | | | | | | 0.002 | | |
| Google Trends "Price of Oil" Query Count (t–3) | | | | | | 0.017* | | |
| Wikipedia "OPEC" Page Views (t–2) | | | | | | | -3.36E-05 | |
| Wikipedia "Price of Oil" Page Views (t–3) | | | | | | | -0.001* | -0.001* |
| GDELT Number of Organizations (t–3) | | | | | | | -0.018 | |
| GDELT Number of Articles (t–2) | | | | | | | -0.004* | -0.004* |
| Number of Observations | 372 | 372 | 372 | 372 | 372 | 372 | 372 | 372 |
| AIC (in-sample) | 1323.31 | 1274.45 | 1230.39 | 1252.68 | 1226.70 | 1219.93 | 1256.15 | 1218.43 |
| BIC (in-sample) | 1354.62 | 1305.43 | 1268.68 | 1283.49 | 1261.16 | 1254.39 | 1290.82 | 1260.55 |
| RMSE (out-of-sample) | 13.878 | 1.887 | 1.810 | 2.001 | 1.752 | 1.970 | 1.833 | 1.683 |
| MAPE (out-of-sample) | 22.579 | 2.898 | 2.762 | 3.044 | 2.674 | 3.016 | 2.794 | 2.498 |

*p<0.05; (t–1) One-day lag; (t–2) Two-day lag; (t–3) Three-day lag.

In terms of out-of-sample evaluation, Google Trends emerges as the best time series predictor for the WTI Crude Oil Price, followed by Twitter. Finally, Model 8 shows that combining Twitter Complexity, Wikipedia "price of oil" page views count and GDELT number of articles returns the best results for both in-sample and out-of-sample predictions. Model 8 graphical performance is presented in Figure 1.

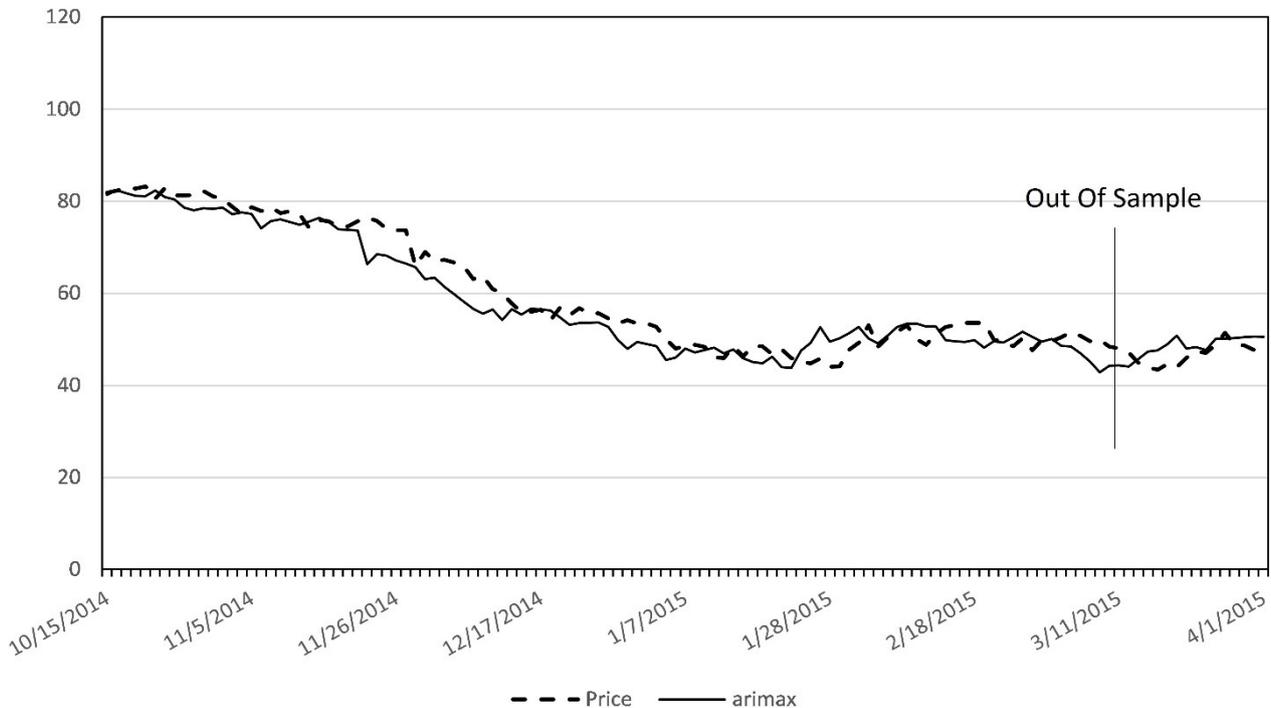

ARIMAX (2,1,4) Prediction

**Figure 1.** Model 8 Prediction.

## 5. Discussion and conclusions

An accurate prediction of the crude oil price can be extremely important, since this price has direct effects on several goods and products and its fluctuations affect the stock markets. Oil prices are not only driven by economic variables, but they are also affected by key events that can create a state of awareness, such as political events, military conflicts, severe climate abnormalities and even big accidents [79]. Global and local awareness is reflected on social media and can be measured tracking indicators like the sentiment of the language used or the frequency of interaction [80]. The global discourse, expressed both online and in newspapers, is linked and partially representative of the events involving the oil industry [81]. In this study, we collected data from four different media platforms (Twitter, Wikipedia, Google Trends and GDELT) and translated them into 10 independent variables, with the intent of making more accurate predictions of the WTI Crude Oil Price. The content that has been analysed was either user generated or extracted from the world's news broadcasts (for the case of GDELT). The informative power of each platform has been tested against the most important predictors coming out from each different media source. The authors also gave evidence to the advantages of integrating information from different platforms, to realize the predictive models. The alternative of just relying on a single platform to operate predictions proved to be less accurate (see Table 4). The results coming from the models show that the social parameters – extracted from the above-mentioned platforms – have remarkably high correlations with the oil price movements. Consistently, significant granger-cause relations were identified for six out of ten predictors. In particular, a higher number of articles on GDELT, as well as more messages related to the oil price on Twitter and a higher number of page views on the Wikipedia pages that were examined, showed a negative relation with the oil price. This is probably due to the fact that the online discourse, as well as the media coverage, is more active when negative events are affecting the oil price. On the other hand, the positive association of the oil price with the Number of Organizations mentioned on GDELT is probably due to the fact that, in several cases, when media articles mention organizations they refer to more favourable topics, such as business agreements or the discovery of new oilfields. Similarly, queries on Google proved to be positively associated with the oil price, although the Google Trends predictors were not significant in the final ARIMAX models. The oil price also showed to be higher when the tweets were more positive and less complex, probably reflecting a more positive sentiment of the online discourse, while participants in the discussion used more shared language. Contrary to expectations, the emotionality of Tweets did not prove to be significantly associated with the oil price (H1c is not supported). This is probably because positive and negative events were clearly distinguishable over time, leading the sentiment to be either positive or negative, without large variations within a single day. The paper results support all the study hypotheses except for H1c.

Using ARIMAX models, the study proved that a combined analysis of Twitter, Wikipedia and GDELT (see Table 4, Model 8) can lead to forecasts for crude oil prices with a reasonably high level of accuracy – thus supporting the advantage of integrating multiple data sources, instead of relying on a single media platform. When considering all the predictors together, a higher price for the crude oil can be forecasted if the number of media articles is lower, as well as the page views on Wikipedia, and if the tweets about the oil price are written in a simpler language. Lastly, it has been proved that the four media platforms can deliver valuable information at different "speeds": Twitter is most informative at a one-day lag, GDELT and Wikipedia at a two-day lag and Google Trends at a three-day lag. From this perspective, the research findings are in contrast with other studies: some scholars claim that Google Trends is the best media source for almost real time forecasting e.g. [42,45].

The main contribution of this study is to propose a new model which can be integrated with the existing crude oil price forecasting techniques. The authors' findings can help decision makers – either firms, private investors, or individuals – when choosing to buy or sell the crude oil, or when making other investments on the stock market. The price of crude oil is not just an asset itself: forecasting it can be extremely useful as it is connected to other stock prices and can have a direct impact on several goods and services, such as transportation costs. Simultaneously, it affects export and import costs, which are part of the Gross Domestic Product. An additional contribution is the combination of social data from four different sources. Previous works, on the other hand, studied a maximum of two platforms to build forecasting models and usually considered each of them individually, e.g. [48,82]. By contrast, this paper implements multiple variable/platform models and explores their effects in time. The aim is to compare different types of available online social platforms, either separated or combined, in order to analyse different online publics, ranging from individuals to governments and news outlets. Finally, the study allows a comparison of the different fore-sighting abilities of each platform, in terms of how many days ahead a platform can predict a price movement before it happens.

This study has several limitations, which the authors are willing to address with future research. The data collected are limited only to the feeds containing the keywords, defined as search terms. Therefore, some other feeds relating to the same subject but using other keywords or expressions might have been missed. For Twitter, the main limitation can be found in its structure. Powerful individuals – who have a big number of followers, a big retweet traffic on their tweets, or are mentioned in a big number of tweets – can lead others to engage in a certain act [83]. Therefore, not all users should have the same weight in calculating either sentiment or activity metrics, and for future research the authors recommend applying different weights to different users. With regard to Wikipedia, the variables have been limited to the page views activity. Future work could consider several other variables, such as the number of users who have contributed to the page, the number of page edits, the time span between different versions of an article, the length and sentiment of each edit, and so on. Concerning Google Trends, the search engine itself is not able to represent all online activities related to the Oil Price. Probably, some traders, hedgers, and speculators skip the search engine, directly referring to specialized platforms. Lastly, regarding the GDELT Project, this study measured the volume of news articles based on keyword search and the count of organizations mentioned in that news. Future works could consider, for instance, the number of mentions, which counts the number of information sources containing mentions of the event.

Generally speaking, several extensions to this work are possible. The approach and the variables proposed in this study could be easily integrated in other existing models, to improve their predictive power. With a step in the direction of predicting economic indicators and stock market prices, authors maintain the importance of extracting relevant knowledge from a large set of public available data sources.

### Notes

1. https://blog.twitter.com/2012/twitter-turns-six
2. https://about.twitter.com/company
3. http://guardian.galaxyadvisors.com/
4. https://cloudplatform.googleblog.com/2014/05/worlds-largest-event-dataset-now-publicly-available-in-google-bigquery.html

### Funding

This research received no specific grant from any funding agency in the public, commercial, or not-for-profit sectors.